\documentclass{ws-ijmpa}
\usepackage{amssymb,graphicx,amsmath,graphics,subfigure,xfrac,url,xcolor}

\begin{document}

\title{The DKP Equation in Presence of a Cusp Potential: Transmission Resonances and Bound States}
\author{Vicente A. Ar\'evalo}
\address{Yachay Tech University, School of Physical Sciences and Nanotechnology, Hda. San Jos\'e S/N y Proyecto Yachay, 100119, Urcuqu\'i, Ecuador.\\
vicente.arevalo@yachaytech.edu.ec}

\author{Sebasti\'an Valladares}
\address{Facultad de Física, Universidad de Sevilla, 41012--Sevilla, España.\\
sebvalsan@alum.us.es}

\author{Clara Rojas}
\address{Yachay Tech University, School of Physical Sciences and Nanotechnology, Hda. San Jos\'e S/N y Proyecto Yachay, 100119, Urcuqu\'i, Ecuador.\\
crojas@yachaytech.edu.ec}

\maketitle

\pub{Received (February 19, 2024)}{Revised (\today)}

\begin{abstract}
In this article, we solve the Duffin--Kemmer--Petiau (DKP) equation in the presence of the cusp potential for spin--one particles. We derived the scattering solutions and calculated the bound states in terms of the Whittaker functions. We show that transmission resonances are present, as well as the particle--anti-particle bound states.

\end{abstract}

\ccode{PACS Nos.: 03.65.Pm, 03.65.Nk, 03.65.Ge}

\section{Introduction}	
In the literature, there is a relativistic equation called Duffi--Kemmer--Petiau (DKP) equation, which describes the dynamics of scalar (spin--0) and vector (spin--1) particles in
analogy with the Dirac equation for spin--half fields\cite{deMontigny2019,Duffin1938,1939}. Besides, it is essential to mention that the equation was not taken into account due to its equivalence with the Klein--Gordon and Proca equations. Despite this, some aspects cannot be explained without using the DKP equation.

The DKP formalism with different types of couplings is used in a wide area of physics, including applications in Quantum Chromodynamics at large and short distances\cite{Gribov1999}, studies of covariant Hamiltonian dynamics\cite{KANATCHIKOV2000107}, symmetries of very special relativity, thermal properties of bosons\cite{aounallah2020thermal}, among others\cite{chetouani:2004,boutabia2005solution}. Therefore, it represents a promising area of research.

Additionally, the DKP equation has been used to model several interactions, including, but not limited to, those involved in many scattering processes in nuclear physics\cite{CHARGUI2023128538}. Furthermore, one--dimensional potentials have been much more prevalent in DKP formalism in recent decades due to the substantial assistance that the resultant simple equations offer for understanding physical systems in higher dimensions\cite{de2016quantum}. Among the potentials used, we can mention the double--step potential\cite{de2012scattering,de2015transmission}, the Varshni potential model\cite{oluwadare2017scattering},  the DKP oscillator \cite{castro2011spinless,kulikov2005alternative}, hyperbolic tangent potential\cite{valladares2023superradiance}, also to study the transmission resonances in an asymmetric cusp potential\cite{Sogut2010}. In this sense, the purpose of this article is to address the behavior of a spinor particle for a cusp potential, including scattering and bound states of the wave equations, which are of great interest in such quantum mechanical systems and it is a well--understood problem for the Schr\"odinger equation\cite{newton2013scattering}.

\begin{figure}[h]
\begin{center}
\includegraphics[scale=0.7]{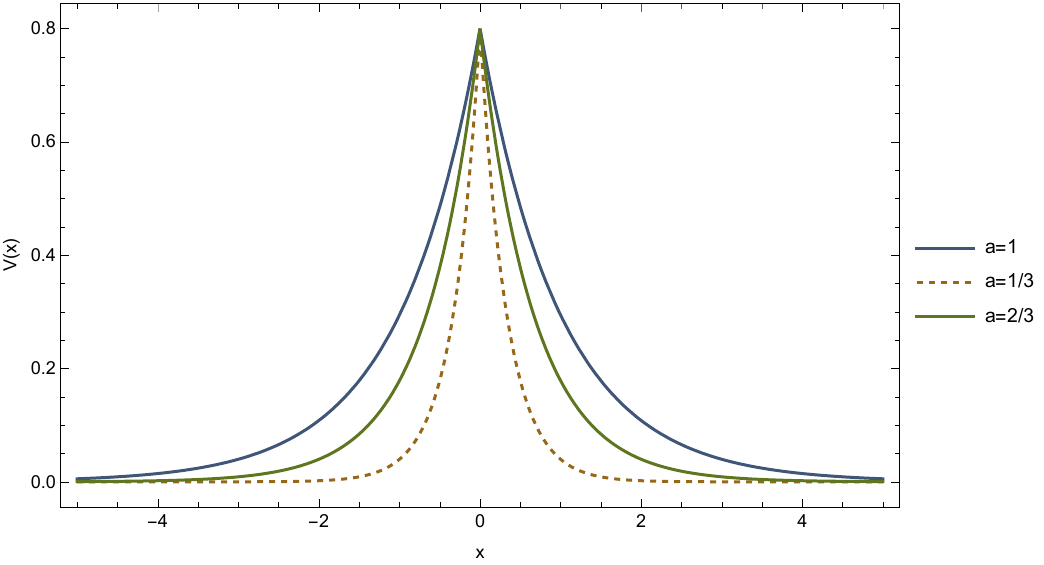}
\caption{Behaviour of a cusp potential barrier vs the spatial coordinate $x$ for different values of the parameter $a$.}
\label{fig:cuspide1}
\end{center}
\end{figure}

The one--dimensional symmetric cusp potential barrier is given by

\begin{equation}
\label{cuspide1}
V(x)= V_{0}\exp\left(-\dfrac{|x|}{a}\right),
\end{equation}
it is an asymptotically vanishing potential for large values of the space variable $x$. The parameter $V_0$ shows the height of the barrier (see Fig. \ref{fig:cuspide1}) or the depth of the well, depending on its sign. The positive constant a determines the shape of the potential\cite{ikot2015scattering}. Moreover, the cusp potential can be regarded as a screened one--dimensional Coulomb potential if $V_0 \leq 0$\cite{villalba2003transmission}.

The paper is organized as follows. In Section \ref{sec_Well}, we solve and analyze the DKP equation for a cusp potential barrier, including the reflection and transmission coefficient. In Section \ref{sec_results}, we obtain and discuss the energy equation that relates the bound states for a cusp potential well. Finally, the conclusions are discussed in Section \ref{sec_conclusion}. We set natural units $\hbar=c=1$, and  $m=1$.


\section{Scattering states}
\label{sec_Well}

As stated above, the Duffin--Kemmer--Petiau (DKP) equation can be used to describe spin--zero and one particles, and with the introduction of an electromagnetic field interaction\cite{chetouani:2004}, it takes the following form  

\begin{equation}
\label{DKP_spin}
\left[i \beta^\mu \left(\partial_\mu+i e A_\mu \right) -1\right]\Psi(\Vec{r},t)=0,
\end{equation}
where $\Psi(\Vec{r},t)$ represents the DKP spinor of the particle in $(3+1)$ dimensions, and the matrices $\beta^{\mu}$ satisfy the following condition $\beta^{\mu}\beta^{\nu}\beta^{\lambda}+\beta^{\lambda}\beta^{\nu}\beta^{\mu}=\eta^{\mu \nu} \beta^{\lambda}+\eta^{\lambda \nu} \beta^{\mu}$ \cite{deMontigny2019,de2010bound}, 

It is important to mention that in $(3+1)$ dimensions, the  $\beta$ matrices are five-dimensional and ten-dimensional for spin--zero and spin--one particles, respectively\cite{boumali2008eigensolutions}. Nevertheless, in $(1+1)$ dimensions, the scalar (spin--zero) and vector (spin--one) sectors are unitarily equivalent\cite{deMontigny2019,lunardi:2017}. It means that the spin--one case has to be treated as a spin--zero case.

As the potential involved is time-independent, the solution of Eq. (\ref{DKP_spin}) can be obtained by $\Psi(x,t)=e^{iEt}\phi(x)$. Moreover, the DKP equation can be then rewritten in $(1+1)$ dimensional representation as follows 

\begin{equation}\label{eq4}
\left\{\beta^{0}\left[E-V(x)\right]+ i\beta^{1}\dfrac{\mathrm{d}}{\mathrm{d}x}-1 \right\}\phi(x)=0,
\end{equation}
where $\phi(x)$ is the spinor, $E$ is the energy, and $V(x)$ represents the coupled potential to the DKP equation. For $(1+1)$ dimension, we consider the following expressions for $\beta^{0}$, and $\beta^{1}$ matrices\cite{lunardi:2017,langueur2021dkp}, with the metric tensor $\eta^{\mu \nu}=\textnormal{diag} \left(1,-1\right)$.
\begin{equation}\label{eqBeta0-1}
    \beta^0 = \left(\begin{array}{ccc}
        0 & 0 & i \\
        0 & 0 & 0 \\
        -i & 0 & 0
    \end{array}\right), \quad
    \beta^1 = \left(\begin{array}{ccc}
        0 & i & 0 \\
        i & 0 & 0 \\
        0 & 0 & 0
    \end{array}\right).
\end{equation}

Since, $\beta^{0}$, and $\beta^{1}$ are $3 \times 3$ matrices, $\phi (x)$ must be decomposed in three components, such that
\begin{equation}
 \phi (x)=\left(\begin{array}{c}\Psi_(x)  \\ \Phi_ (x)\\ \Theta (x)\end{array}\right),
 \label{eqBeta3}
\end{equation}
generally,  $\Psi(x)$,  $\Phi(x)$, $\Theta(x)$ are functions that describe particles in the DKP theory.

From here, it is not difficult to verify that by using Eqs. \eqref{eqBeta3}, and \eqref{eqBeta0-1}, on Eq. (\ref{eq4}) lead to a system of three equations\cite{valladares2023superradiance}.

\begin{equation}
\label{eq7}
\begin{cases}
\left\{\dfrac{\mathrm{d}^2}{\mathrm{d}x^2}+ \left[E-V(x)\right]^2-1\right\}\Psi (x)=0,\vspace{1mm}\\
\Phi(x)=-\dfrac{\mathrm{d}}{\mathrm{d}x}\Psi (x)\\
\Theta (x)=-i\left[E-V(x)\right]\Psi (x)\vspace{1mm}.
\end{cases}
\end{equation}
where it is clear that $\Phi(x)$ and $\Theta(x)$ depend on the Klein--Gordon equation solution (i.e., $\Psi(x)$).

\subsection{Transmission Resonances}

To calculate the transmission resonances we solve Eq. \eqref{eq7} with the cusp potential barrier, which is given by\cite{villalba2003transmission,villalba2006bound}

\begin{equation}
\label{barrier}
V(x)=\left\{\begin{array}{ccc}
\hspace{-0.2cm} V_0 e^{\sfrac{x}{a}} & \text { for } \quad x<0, \\
\hspace{0.1cm} V_0 e^{-\sfrac{x}{a}} & \text { for } \quad x>0.
\end{array}\right.
\end{equation}

The form of this barrier is represented in Fig. \ref{fig:cuspide1}.

To solve the problem, it is important to consider the Klein--Gordon equation for both regions, the positive and negative. For $x<0$. We solve the differential equation

\begin{equation}
\label{left}
\dfrac{\mathrm{d}^2 \phi_L(x)}{\mathrm{d} x^2}+\left[\left(E-V_0 e^{\sfrac{x}{a}}\right)^2-1\right] \phi_L(x)=0 .
\end{equation}

Making the change of variable $y=2 i a V_0 e^{\sfrac{x}{a}}$, 

\begin{equation}
y \dfrac{\mathrm{d}}{\mathrm{d} y}\left(y \dfrac{\mathrm{d} \phi_L}{\mathrm{d} y}\right)-\left[\left(i a E-\dfrac{y}{2}\right)^2+a^2\right] \phi_L=0.    
\end{equation}

Setting $\phi_L=y^{-\sfrac{1}{2}} f(y)$, the previous equation reduces to the Whittaker differential equation.

\begin{equation}
\label{ode_1}
\dfrac{\mathrm{d}^2 f(y)}{\mathrm{d} y^2}+\left[-\dfrac{1}{4}+\dfrac{i a E}{y}+\dfrac{\sfrac{1}{4}-\mu^2}{y^2}\right] f(y)=0.
\end{equation}

Eq. \eqref{ode_1} is the Whittaker differential equation\cite{abramowitz:1965} whose general solution can  be written as
\begin{equation}
\label{sol_1}
\phi_L(y)=c_1 y^{-\sfrac{1}{2}} M_{\kappa, \mu}(y)+c_2 y^{-\sfrac{1}{2}} M_{\kappa,-\mu}(y),
\end{equation}
where $M_{\kappa, \mu}(y)$ is the Whittaker function, and

\begin{equation}
\kappa=i a E, \quad \mu=i a \sqrt{E^2-1}.
\end{equation}

Notice that the general solution given in Eq. (\ref{sol_1}) is made by the incident and the reflected part of the wave. Moreover, the goal is to solve the system of equations Eq. (\ref{eq7}). Therefore, let's analyze both cases separately. 

\subsection{Incident Part}

The solution of the system of the system Eq. (\ref{eq7}) for the incident part is composed by

\begin{align}
\Psi_{inc}(x)&= c_1\dfrac{M_{\kappa, \mu}\left(2iaV_0e^{\sfrac{x}{a}}\right)}{\sqrt{2iaV_0e^{\sfrac{x}{a}}}},\\
\nonumber
\Phi_{inc}(x)&= c_1\dfrac{1 }{2  a  \sqrt{2i a V_0 e^{\sfrac{x}{a}}}}\left[\left(2 \kappa+1-2 i a V_0 e^{\sfrac{x}{a}}\right) M_{\kappa,\mu}\left(2 i a e^{\sfrac{x}{a}}V_0\right)\right.\\
&-\left.\left(2 \kappa+2 \mu+1\right) M_{\kappa+1,\mu}\left(2 i a e^{\sfrac{x}{a}} V_0\right)\right],\\
\Theta_{inc}(x)&=i~c_1\dfrac{\left(e^{\sfrac{x}{a}}V_0-E\right)M_{\kappa, \mu}\left(2iaV_0e^{\sfrac{x}{a}}\right)}{\sqrt{2iaV_0e^{\sfrac{x}{a}}}}.
\end{align}

For simplicity, we can group these solutions in a vector representation

\begin{equation}
\label{Gi}
G_{inc}=\begin{pmatrix} \quad
c_1\dfrac{M_{\kappa, \mu}\left(2iaV_0e^{\sfrac{x}{a}}\right)}{\sqrt{2iaV_0e^{\sfrac{x}{a}}}} \\ 
c_1\dfrac{1 }{2  a  \sqrt{2i a V_0 e^{\sfrac{x}{a}}}}\left[\left(2 \kappa+1-2 i a V_0 e^{\sfrac{x}{a}}\right) M_{\kappa,\mu}\left(2 i a e^{\sfrac{x}{a}}V_0\right)\right.\\
\hspace{2.1cm}-\left.\left(2 \kappa+2 \mu+1\right) M_{\kappa+1,\mu}\left(2 i a e^{\sfrac{x}{a}} V_0\right)\right]\\
i~c_1\dfrac{\left(e^{\sfrac{x}{a}}V_0-E\right)M_{\kappa, \mu}(2iaV_0e^{\sfrac{x}{a}})}{\sqrt{2iaV_0e^{\sfrac{x}{a}}}}

\end{pmatrix},
\end{equation}
where $G_{inc}$ represents the vector that contains the incident solutions.

\subsection{Reflected Part}

Similar to before, the solution for the reflected part is composed by:

\begin{eqnarray}
\Psi_{ref}(x) &=& c_2\dfrac{  M_{\kappa, -\mu}(2iaV_0e^{\sfrac{x}{a}})}{\sqrt{2iaV_0e^{\sfrac{x}{a}}}},\\
\nonumber
\Phi_{ref}(x) &=&c_2 \dfrac{1}{2  a  \sqrt{2i a V_0 e^{\sfrac{x}{a}}}} \left[\left(2 \kappa+1-2 i a V_0 e^{\sfrac{x}{a}}\right) M_{\kappa,-\mu}\left(2 i a e^{\sfrac{x}{a}} V_0\right)\right.\\
&+&\left.\left(2 \mu-1-2 \kappa\right) M_{\kappa+1,-\mu}\left(2 i a e^{\sfrac{x}{a}} V_0\right)\right],\\
\Theta_{ref}(x) &=& i~c_2\dfrac{(e^{\sfrac{x}{a}}V_0-E)M_{\kappa, -\mu}\left(2iaV_0e^{\sfrac{x}{a}}\right)}{\sqrt{2iaV_0e^{\sfrac{x}{a}}}},
\end{eqnarray}

\newpage
Or in vector representation:

\begin{equation}\label{Gr}
G_{ref}=\begin{pmatrix} \quad
c_2\dfrac{  M_{\kappa, -\mu}\left(2iaV_0e^{\sfrac{x}{a}}\right)}{\sqrt{2iaV_0e^{\sfrac{x}{a}}}}   \\
c_2\dfrac{1 }{2  a  \sqrt{2i a V_0 e^{\sfrac{x}{a}}}} \left[\left(2 \kappa+1-2 i a V_0 e^{\sfrac{x}{a}}\right) M_{\kappa,-\mu}\left(2 i a e^{\sfrac{x}{a}} V_0\right)\right.\\
\hspace{2cm}+\left.\left(2 \mu-1-2 \kappa\right) M_{\kappa+1,-\mu}\left(2 i a e^{\sfrac{x}{a}} V_0\right)\right]\\
i~c_2\dfrac{\left(e^{\sfrac{x}{a}}V_0-E\right)M_{\kappa, -\mu}\left(2iaV_0e^{\sfrac{x}{a}}\right)}{\sqrt{2iaV_0e^{\sfrac{x}{a}}}}
\end{pmatrix},
\end{equation}
where $G_{ref}$ is the vector that contains the reflected solutions.

Now we consider the solution for $x>0$. In this case, the differential equation to solve is

\begin{equation}
\label{ode_2}
\dfrac{\mathrm{d}^2 \phi_R(x)}{\mathrm{d} x^2}+\left[\left(E-V_0 e^{-\sfrac{x}{a}}\right)^2-1\right] \phi_R(x)=0.
\end{equation}

Making the variable change $z=2 i a V_0 e^{-\sfrac{x}{a}}$, Eq. \eqref{ode_2} can be written as

\begin{equation}
z \dfrac{\mathrm{d}}{\mathrm{d} z}\left[z \dfrac{\phi_R(z)}{\mathrm{d} z}\right]-\left[\left(i a E-\dfrac{z}{2}\right)^2+a^2\right] \phi_R(z)=0.
\end{equation}

Putting $\phi_R(z)=z^{-\sfrac{1}{2}} g(z)$ we obtain the Whittaker differential equation\cite{abramowitz:1965}

\begin{equation}
\dfrac{\mathrm{d}^2 g(z)}{\mathrm{d} z^2}+\left[-\dfrac{1}{4}+\dfrac{i a E}{z}+\dfrac{\sfrac{1}{4}-\mu^2}{z^2}\right] g(z)=0,
\end{equation}
which solution is given by\cite{abramowitz:1965}

\begin{equation}
\label{sol_2}
\phi_R(z)=c_3 z^{-\sfrac{1}{2}} M_{\kappa,-\mu}(z).
\end{equation}

The general solution given in Eq. (\ref{sol_2}) corresponds to the transmitted part of the wave. 

\subsection{Transmitted part}

The solutions for the right side are,

\begin{eqnarray}
\Psi_{trans}(x)&=&c_3 \dfrac{M_{\kappa,-\mu}\left(2iaV_0e^{-\sfrac{x}{a}}\right)}{\sqrt{2iaV_0e^{-\sfrac{x}{a}}}},\\
\nonumber
\Phi_{trans}(x)&=&c_3\dfrac{\sqrt{iaV_0e^{-\sfrac{x}{a}}}}{\sqrt{2}a^2V_0}\left[\left(ie^{\sfrac{x}{a}}\left(1+2\kappa\right)+2aV_0\right) M_{\kappa,-\mu}(2iaV_0e^{-\sfrac{x}{a}})\right.\\
&-&\left. i~e^{\sfrac{x}{a}}\left(1+2\kappa-2\mu\right) M_{\kappa+1,-\mu}\left(2iaV_0e^{-\sfrac{x}{a}}\right) \right],\\
\Theta_{trans}(x)&=&i~c_3\dfrac{\left(e^{-\sfrac{x}{a}}V_0-E\right)M_{\kappa,-\mu}(2iaV_0e^{-\sfrac{x}{a}})}{\sqrt{2iaV_0e^{-\sfrac{x}{a}}}},
\end{eqnarray}

In vector representation,

\begin{equation}
\label{Gt}
 G_{trans}=\begin{pmatrix}  
\dfrac{c_3 M_{\kappa,-\mu}\left(2iaV_0e^{-\sfrac{x}{a}}\right)}{\sqrt{2iaV_0e^{-\sfrac{x}{a}}}}  \\ 
c_3\dfrac{\sqrt{iaV_0e^{-\sfrac{x}{a}}}}{\sqrt{2}a^2V_0}\left[\left(i~e^{\sfrac{x}{a}}\left(1+2\kappa\right)+2aV_0\right) M_{\kappa,-\mu}\left(2iaV_0e^{-\sfrac{x}{a}}\right)\right.\\
\hspace{2cm}\left. -i~e^{\sfrac{x}{a}}\left(1+2\kappa-2\mu\right) M_{\kappa+1,-\mu}\left(2iaV_0e^{-\sfrac{x}{a}}\right) \right]  \\
\dfrac{i~c_3\left(e^{-\sfrac{x}{a}}V_0-E\right)M_{\kappa,-\mu}\left(2iaV_0e^{-\sfrac{x}{a}}\right)}{\sqrt{2iaV_0e^{-\sfrac{x}{a}}}} 
\end{pmatrix},
\end{equation}
where $G_{trans}$ is the vector that contains the transmitted solutions.

\subsection{Asymptotic behavior}\label{asi}

Once the solutions of the DKP equation have been obtained in the analytic form. Now, we analyze the scattering states for the cusp potential considering that the asymptotic behavior of the Whittaker functions\cite{abramowitz:1965} is given by 

\begin{equation}
\label{condition}
M_{\kappa,\mu}(y)=e^{\sfrac{-y}{2}}y^{\frac{1}{2}+\mu},
\end{equation}
when $y\rightarrow 0$.

In the same way considering the incident solution \eqref{Gi}, when $x\rightarrow -\infty$, we have the following result

\begin{equation}
\label{ai}
G_{inc} \longrightarrow c_1\left(2 i a V_0\right)^\mu e^{i\sqrt{E^2-1} x }\left(\begin{array}{c}
1 \\
\dfrac{-\mu}{ a}\\
-iE 
\end{array}\right),
\end{equation}
which behaves like a plane wave that is traveling to the right side.

Now, for the reflection part \eqref{Gr}, we have the following expression

\begin{equation}
\label{ar_1}
G_{ref} \longrightarrow c_2\left(2 i a V_0\right)^{-\mu} e^{-i\sqrt{E^2-1} x }\left(\begin{array}{c}
1 \\
\dfrac{\mu}{ a}\\
 -iE
\end{array}\right) ,
\end{equation}
which, as in the previous case, represents a plane wave, but in this case, it is traveling to the left side.

Finally, considering the right solution \eqref{Gt} when $x\rightarrow \infty$, we have 

\begin{equation}
\label{ar_2}
G_{trans} \longrightarrow c_3\left(2 i a V_0\right)^{-\mu} e^{i\sqrt{E^2-1} x }\left(\begin{array}{c}
1 \\
\dfrac{-\mu}{ a} \\
-iE
\end{array}\right) ,
\end{equation}

From the previous results, it is possible to define the reflection and transmission coefficients in terms of the normalization constants as follows

\begin{align}
R&=\left|\dfrac{c_2\left(2 i a V_0\right)^{-\mu}}{c_1\left(2 i a V_0\right)^{\mu}}\right|^2\label{r},\\ 
T&=\left|\dfrac{c_3\left(2 i a V_0\right)^{-\mu}}{c_1\left(2 i a V_0\right)^{\mu}}\right|^2\label{t}.
\end{align}

\subsection{Reflection and Transmission coefficients}

In the previous subsection \ref{asi}, we got expressions for the reflection and transmission coefficients, which depend on the constants $c_1$, $c_2$, and~$c_3$ that need to be calculated.

To calculate the required constants, the next step is to form a general solution for the DKP system Eq. (\ref{eq7}) and study the continuity at $x=0$. For it, let us use the matrices of Eqs. (\ref{Gi}), (\ref{Gr}), and (\ref{Gt}).

\begin{equation}
\label{Gnew}
G_{inc}(x=0)+G_{ref}(x=0)=G_{trans}(x=0).
\end{equation}

After applying the condition (\ref{Gnew}), it is possible to have the following system of equations. 
\begin{eqnarray}
\label{sys_eq1}
eq_1&=&\left(c_2-c_3\right)M_{\kappa,-\mu}\left(2iaV_0\right)+c_1M_{\kappa,\mu}\left(2iaV_0\right)=0,\\
\label{sys_eq2}
eq_2&=&iV_0
\left\{ \dfrac{ia \left[ \left(c_2+c_3\right)\left(\left(1-2\mu\right)^2-4\kappa^2\right) M_{\kappa,1-\mu}\left(2iaV_0\right)\right]}{\left(1-2\mu\right)^2\left(\mu-1\right)} \right. \\
\nonumber
\quad &+&\dfrac{4\left(c_2+c_3\right)\left[\mu\left(2\mu-1\right)-2ia\kappa V_0\right] M_{\kappa,-\mu}\left(2iaV_0\right)}{2\mu-1}\\
\nonumber
\quad &+&\dfrac{c_1}{\left(1+2\mu\right)^2}\left[-4\left(1+2\mu\right)\left(2\mu^2+\mu-2ia\kappa V_0\right) M_{\kappa,\mu}\left(2iaV_0\right) \right.\\
\nonumber
\quad &-&\left. \dfrac{ia V_0 \left[-4\kappa^2+\left(1+2\mu\right)^2\right] M_{\kappa,\mu+1}\left(2iaV_0\right)}{1+\mu}\right\}=0,\\
\label{sys_eq3}
eq_3&=&(V_0-E) (c_1M_{\kappa,\mu}(2 i a V_0)+(c_2-c_3) M_{\kappa,-\mu}(2 i a V_0))=0.
\end{eqnarray}

Finally, to determine the reflection and transmission coefficients, it is necessary to calculate the value of the coefficients $c_2$ and $c_3$. By solving the system of equations, Eqs. (\ref{sys_eq1})--(\ref{sys_eq3}) for $c_2$ and $c_3$, it is not difficult to verify that they take the following values.

\begin{align}
c_2&=\frac{1}{16} c_1\left\{\left[  8 (1-2 \mu )^2 (\mu -1) \left(a V_0 \left(4 \kappa ^2-(2 \mu +1)^2\right)
  M_{\kappa ,\mu +1}\left(2 i a V_0\right)\right.\right.\right.\\ \nonumber&+\left.\left.\left.4 (\mu +1) (2 \mu +1) \left(2 a \kappa  V_0+i \mu  (2 \mu +1)\right)
   M_{\kappa ,\mu }\left(2 i a V_0\right)\right)\right]\right.\\ \nonumber &\big / \left. \left[ (\mu +1) (2 \mu +1)^2 \left(a V_0 \left(4 \kappa ^2-(1-2 \mu
   )^2\right) M_{\kappa ,1-\mu }\left(2 i a V_0\right)\right.\right.\right.\\ \nonumber&+\left.\left.\left.4 (\mu -1) (2 \mu -1) \left(2 a \kappa  V_0+i \mu  (2 \mu
   -1)\right) M_{\kappa ,-\mu }\left(2 i a V_0\right)\right) \right]\right.\\ \nonumber&- \left.\frac{8 M_{\kappa ,\mu }\left(2 i a
   V_0\right)}{M_{\kappa ,-\mu }\left(2 i a V_0\right)} \right\}
\end{align}
\begin{align}
c_3&=\frac{1}{2} c_1\left\{\left[  (\mu -1) (1-2 \mu )^2 \left(4 (\mu +1) (2 \mu +1) \left(-2 i a \kappa  V_0+2 \mu
   ^2+\mu \right) \right.\right.\right.\\ \nonumber & \times  M_{\kappa ,\mu }\left(2 i a V_0\right)-\left.\left. \left. i a V_0 \left(4 \kappa ^2-(2 \mu +1)^2\right) M_{\kappa
   ,\mu +1}\left(2 i a V_0\right)\right) \right]\right.\\ \nonumber & \big / \left. \left[(\mu +1) (2 \mu +1)^2 \left(4 (\mu -1) (2 \mu -1) \left(\mu  (2 \mu -1)-2
   i a \kappa  V_0\right) M_{\kappa ,-\mu }\left(2 i a V_0\right)\right. \right.\right.\\ \nonumber
   &-\left.\left.\left. i a V_0 \left(4 \kappa ^2-4 \mu ^2+4 \mu -1\right)
   M_{\kappa ,1-\mu }\left(2 i a V_0\right)\right)\right]\right.\\ \nonumber &+ \left.\frac{M_{\kappa ,\mu }\left(2 i a V_0\right)}{M_{\kappa ,-\mu
   }\left(2 i a V_0\right)} \right\}
\end{align}
Finally, we can obtain the expressions for the reflection coefficient $R$ and the transmission coefficient replacing the values of $c_2$ and $c_3$ in the formulas \eqref{r} and \eqref{t}.

\begin{figure}[th]
\begin{center}
\includegraphics[scale=0.7]{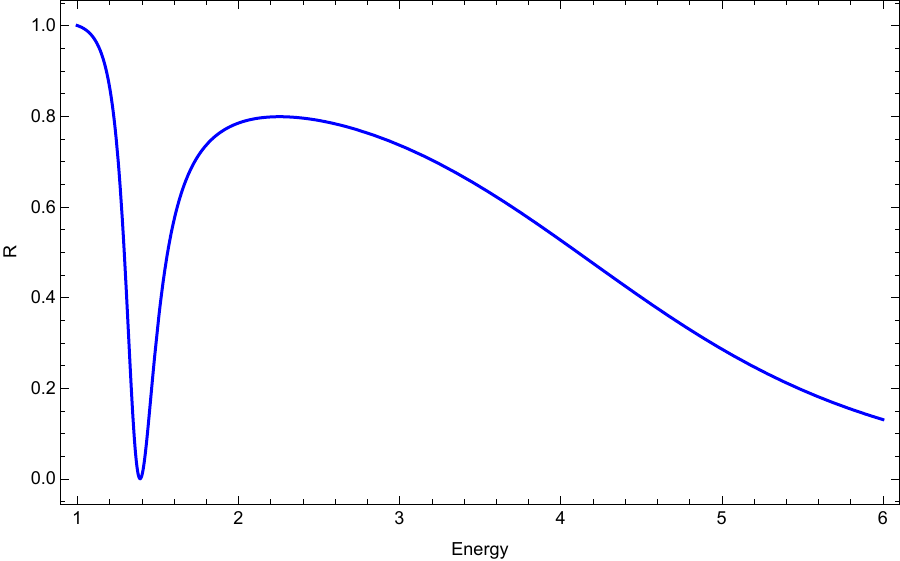}
\caption{The reflection coefficient $R$ (blue line) varying energy $E$ for the cusp potential using $a=0.6$, and $V_0=4$.}
\label{fig_R}
\end{center}
\end{figure}

\begin{figure}[th]
\begin{center}
\includegraphics[scale=0.7]{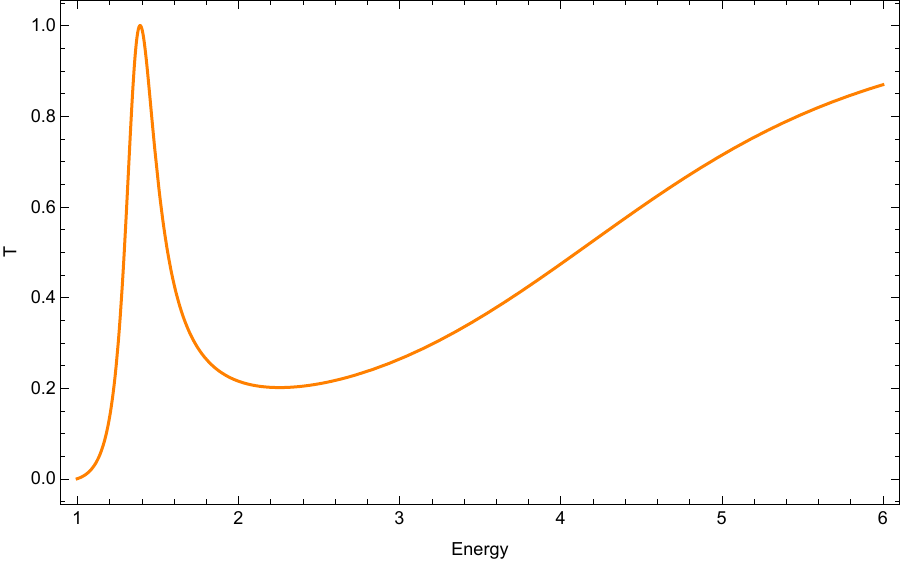}
\caption{The transmission coefficient $T$ (orange line) varying energy $E$ for the cusp potential using $a=0.6$, and $V_0=4$.}
\label{fig_T}
\end{center}
\end{figure}

\begin{figure}[th]
\begin{center}
\includegraphics[scale=0.7]{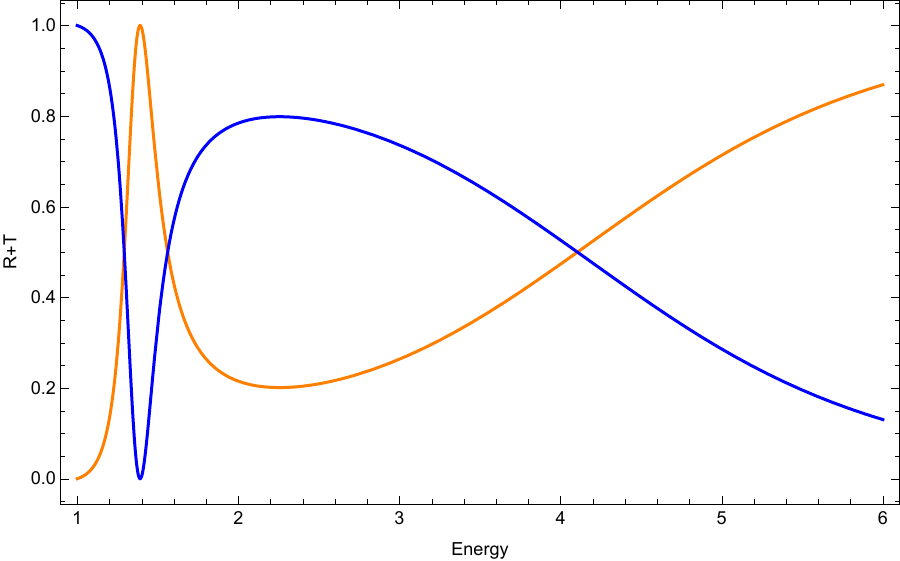}
\caption{The reflection $R$ (blue line) and transmission $T$ (orange line) coefficients varying energy $E$ for the cusp potential using $a=0.6$,~and~ $V_0=4$.}
\label{fig_RT}
\end{center}
\end{figure}

The plot for each coefficient can be observed in Figs.  \ref{fig_R}, and  \ref{fig_T}. A plot with both coefficients is also provided in Fig. \ref{fig_RT}. In which the behavior of the $R$ and $T$ coefficient can be studied. It is clear that the unitary condition $R+T=1$ is accomplished in all the energy spectrum. 

Furthermore, looking at Fig. \ref{fig_T}, it is possible to identify a resonance peak occurring around $E\approx 1.5$. it implies that the wave solution will be able to tunnel the potential barrier without problems.

\begin{figure}[th]
\begin{center}
\includegraphics[scale=0.7]{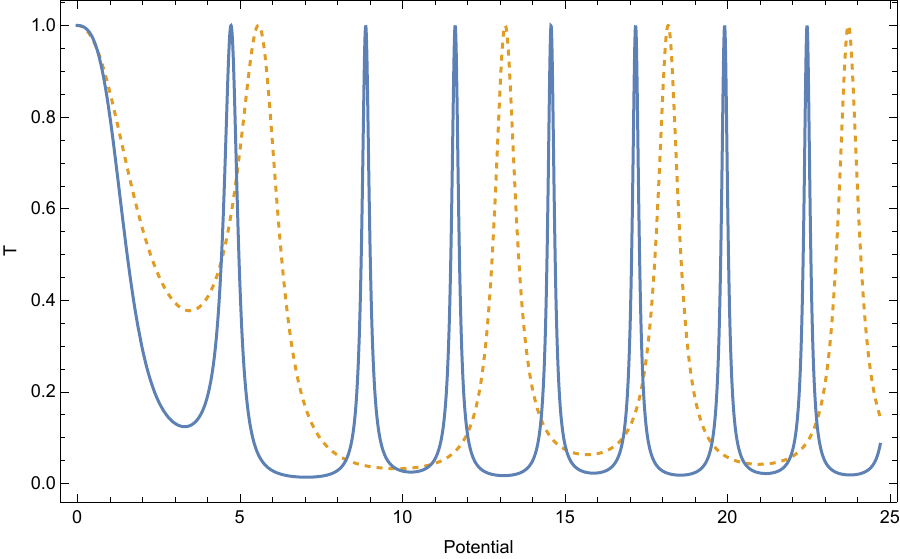}
\caption{Transmission $T$ coefficients vs. $V$ for the cusp potential using $E=2$. Here the parameters are $a=1/3$~(Dashed line), ~$a=2/3$~ (Blue line).}
\label{fig_po}
\end{center}
\end{figure}

\vspace{10mm}
Additionally, we have plotted the transmission coefficient dependence on the strength of the cusp potential for different shape parameters $a$, Fig. \ref{fig_po}. We have noticed that after a particular value of the potential, the transmission resonances occur periodically in steps of $\Delta V \approx 2.93$ for $a=\sfrac{2}{3}$~(Blue line) and~ $\Delta V \approx 5.01$ for $a=\sfrac{2}{3}$~(Dashed line). The difference in the resonances happens due to the shape of the potential; when the potential is broad, the resonances occur much faster than in the narrow case because, for the incoming wave, a broad potential represents less difficulty in tunneling the barrier than the other case.

\section{Bound States}	
\label{sec_results}

The quantum mechanical nature of the bound states arises from the possibility that a wave state can be confined in a potential\cite{hsu2016bound,cohen1986quantum}.
Additionally, a potential will support at least a bound state if the following condition is satisfied\cite{buell1995potentials},

\begin{equation}
\label{boundstate}
\int_{-\infty}^{+\infty}~V\left(x\right)\mathrm{d}x~< 0.
\end{equation}

Now, using condition \eqref{boundstate} for the potential \eqref{cuspide1} with negative $V_{0}$ it is possible to find that

\begin{equation}
\label{boundstate1}
\int_{-\infty}^{+\infty}~V\left(x\right)dx=~-2aV_{0}~< 0,
\end{equation}
which implies that independently of the potential parameters $a$~and~$V_{0}$, the cusp potential well will contain a bound state solution.

As a consequence of the previous result, in this section, we are interested in studying the bound state solution for the DKP equation in the presence of a cusp potential  well given by the expression,

\begin{equation}
\label{well}
V(x)=\left\{\begin{array}{ccc}
\hspace{-0.15cm} -V_0 e^{\sfrac{x}{a}} & \text {for} \quad x<0,\\
-V_0 e^{-\sfrac{x}{a}} & \text {for} \quad x>0.
\end{array}\right.
\end{equation}

To solve the problem, based on Eq. \eqref{eq7}, it is important to consider the Klein--Gordon equation for both regions, positive and negative. For $x<0$, we solve the differential equation

\begin{equation}
\label{left_2}
\dfrac{\mathrm{d}^2 \phi_L(x)}{\mathrm{d} x^2}+\left[\left(E+V_0 e^{\sfrac{x}{a}}\right)^2-1\right] \phi_L(x)=0.
\end{equation}

Considering the change of variable $y=2 i a V_0 e^{\sfrac{x}{a}}$, Eq. \ref{left_2} becomes

\begin{equation}
y \dfrac{\mathrm{d}}{\mathrm{d} y}\left[y \dfrac{\phi_L(y)}{\mathrm{d} y}\right]-\left[\left(i a E+\dfrac{y}{2}\right)^2+a^2\right] \phi_L(y)=0 .
\end{equation}

Taking into account the following solution $\phi_L(y)=y^{-\sfrac{1}{2}} f(y)$  it is possible to obtain the Whittaker differential equation

\begin{equation}
\label{w}
\dfrac{\mathrm{d}^2f(y)}{\mathrm{d} y ^2}+\left[-\dfrac{1}{4}-\dfrac{i a E}{y}+\dfrac{1 / 4-a^2\left(1-E^2\right)}{y^2}\right] f(y)=0 
\end{equation}

The general solution of Eq. \eqref{w} can be expressed in terms of Whittaker functions $M_{\kappa, \mu}(y)$ and $W_{\kappa, \mu}(y)$ as\cite{abramowitz:1965}

\begin{equation}
\phi_L(y)=c\, y^{-\sfrac{1}{2}} M_{\kappa \mu}(y)+d\, y^{-\sfrac{1}{2}} W_{\kappa \mu}(y),
\end{equation}
where $c$ and $d$ are arbitrary constants. Moreover, $\kappa$ and $\mu$ are defined as follows

$$
\kappa=-i a E, \quad \mu=a \sqrt{1-E^2} .
$$

It is necessary to follow the same procedure to obtain the solution for $x>0$.
In this case, we have the differential equation is given by

\begin{equation}
\label{right}
\dfrac{\mathrm{d}^2 \phi_R(x)}{\mathrm{d} x^2}+\left[\left(E+V_0 e^{-\sfrac{x}{a}}\right)^2-1\right] \phi_R(x)=0 
\end{equation}

With the change of variable $z=2 i a V_0 e^{-\sfrac{x}{a}}$, Eq. \eqref{right}  becomes

\begin{equation}
z \dfrac{\mathrm{d}}{\mathrm{d} z}\left[z \dfrac{\phi_R(z)}{\mathrm{d} z}\right]-\left[\left(i a E+\dfrac{z}{2}\right)^2+a^2\right] \phi_R(z)=0 .
\end{equation}

Considering $\phi_R(z)=z^{-\sfrac{1}{2}} g(z)$ we obtain the differential equation

\begin{equation}
\dfrac{\mathrm{d}^2 g(z)}{\mathrm{d} z^2}+\left[-\dfrac{1}{4}-\dfrac{i a E}{z}+\dfrac{\sfrac{1}{4}-a^2\left(1-E^2\right)}{z^2}\right] g(z)=0,
\end{equation}
whose general solution is\cite{abramowitz:1965}

\begin{equation}
\phi_R(z)=e\,  z^{-\sfrac{1}{2}} M_{\kappa \mu}(z)+f z^{-\sfrac{1}{2}} W_{\kappa \mu}(z),
\end{equation}
where $e$ and $f$ are arbitrary constants.

Since we are looking for bound states with the potential \eqref{well}, we choose to work with regular solutions $\phi_L(y)$ and $\phi_R(z)$ along the $x$ axis\cite{villalba2006bound},

\begin{eqnarray}
\phi_L(y)&=&c y^{-\sfrac{1}{2}} M_{\kappa, \mu}(y), \\
\phi_R(z)&=&e z^{-\sfrac{1}{2}} M_{\kappa, \mu}(z) .
\end{eqnarray}

Once we have the solutions related to the Klein--Gordon equation, it is possible to calculate the components of the wave spinor of the DKP equation. The components are expressed in Eq. \eqref{eq7}.

Based on equation \eqref{eqBeta3} it is possible to get the following results for $x<0$,

\begin{equation}
\begin{aligned}
& \phi_L(x)=\left(\begin{array}{l}
\Psi_L(x) \\
\Phi_L(x) \\
\Theta_L(x)
\end{array}\right),
\end{aligned}
\end{equation}

Where the components of the spinor are expressed as follows

\begin{eqnarray}
\Psi_L(x)&=&c_1\left(2 i a V_0 e^{\sfrac{x}{a}}\right)^{-\sfrac{1}{2}} M_{k,\mu}\left(2 i a V_0 e^{\sfrac{x}{a}}\right), \\
\nonumber
\Phi_L(x)&=&c_1\dfrac{1}{2 a}\left(2 i a V_0 e^{\sfrac{x}{a}}\right)^{-\sfrac{1}{2}}\Big[\left(1+2 k-2 i a V_0 e^{x l a}\right) M _{k, \mu}\left(2 ia V_0 e^{\sfrac{x}{a}}\right)\\
&-&(1+2 k+2 \mu) M_{1+k}, \mu\left(2 i a V_0 e^{\sfrac{x}{a}}\right)\Big]\\
\Theta_L(x)&=&- i~c_1 \left(E+ V_0e^{\sfrac{x}{a}}\right)\left(2 i a V_0 e^{\sfrac{x}{a}}\right)^{-\sfrac{1}{2}} M_{k, \mu}\left(2 i a V_0 e^{\sfrac{x}{a}}\right). 
\end{eqnarray}

\bigskip
For $x>0$, the same procedure is applied to get the wave spinor

\begin{equation}
\begin{aligned}
& \phi_R(x)=\left(\begin{array}{l}
\Psi_R(x) \\
\Phi_R(x) \\
\Theta_R(x)
\end{array}\right),
\end{aligned}
\end{equation}
where the components of the spinor are expressed as follows

\begin{eqnarray}
\Psi_R(x)&=& c_2\left(2 i a V_0 e^{-\sfrac{x}{a}}\right)^{-\sfrac{1}{2}} M_{k,\mu} \left(2 i a V_0 e^{-\sfrac{x}{a}}\right),  \\
\nonumber
\Phi_R(x)&=&c_2\dfrac{ \sqrt{i a V_0 e^{-\sfrac{x}{a}}}}{2 \sqrt{2} a^2  V_0}\Big\{\left[i e^{\sfrac{x}{a}}(1+2 k)+2  a V_0\right] M_{k, \mu}\left(2 i a V_0 e^{-\sfrac{x}{a}}\right)\\
 &-&i e^{\sfrac{x}{a}}\left(1+2 k+2 \mu\right) M_{1+k}, \mu\left(2 i a V_0 e^{-\sfrac{x}{a}}\right)\Big\},\\
\Theta_R(x)&=&-\frac{c_2}{\sqrt{2} a V_0} \left(e^{\sfrac{x}{a}} E+V_0 \right) \sqrt{i a V_0 e^{-\sfrac{x}{a}}}M_{k,\mu}\left(2 i a V_0 e^{-\sfrac{x}{a}}\right )^{2}.
\end{eqnarray}

To find the bound states, it is important to consider $|E|<1$, which can be found by requiring continuity of the wave function at $x = 0$ and square--integrability of the wave
function over all space. Then continuity of the spinor at $x = 0$\cite{Sogut2010}.

\begin{equation}
\label{cont}
\phi_L(x=0)=\phi_R(x=0).
\end{equation}

Applying the condition of Eq. \eqref{cont}, it is possible to have the present results for the left side.

\begin{eqnarray}
\Psi_L(0)&=&c_1\left(2 i a V_0 \right)^{-\sfrac{1}{2}} M_{k,\mu}\left(2 i a V_0 \right), \\
\nonumber
\Phi_L(0)&=&\dfrac{ i c_1 V_0}{2 \sqrt{2} (i a V_0)^{3/2}}\Big\{\left[1+2 k-2 i a V_0\right] M_{k, \mu}\left(2 i a V_0 \right)\\
 &-& \left(1+2 k+2 \mu\right) M_{1+k}, \mu\left(2 i a V_0 \right)\Big\}\\
\Theta_L(0)&=&\frac{a c_1 V_0}{\sqrt{2} (i a V_0)^{3/2}} \left( E+V_0 \right) M_{k,\mu}\left(2 i a V_0 \right )^{2}.  
\end{eqnarray}

And now for the right side

\begin{eqnarray}
\Psi_R(0)&=& c_2\left(2 i a V_0 \right)^{-\sfrac{1}{2}} M_{k,\mu} \left(2 i a V_0 \right),  \\
\nonumber
\Phi_R(0)&=&\dfrac{ i c_2 V_0}{2 \sqrt{2} (i a V_0)^{3/2}}\Big\{\left[-1-2 k+2 i a V_0\right] M_{k, \mu}\left(2 i a V_0 \right)\\
 &+& \left(1+2 k+2 \mu\right) M_{1+k}, \mu\left(2 i a V_0 \right)\Big\},\\
\Theta_R(0)&=&\frac{a c_2 V_0}{\sqrt{2} (i a V_0)^{3/2}} \left( E+V_0 \right) M_{k,\mu}\left(2 i a V_0 \right )^{2}.
\end{eqnarray}


Now, equalling each component of the equations, it is possible to have the following relations

\begin{equation}
c_1=c_2.
\end{equation}

This leads to the following algebraic equation to find the values of the energy $E$

\begin{equation}
\label{energy}
\left(1+2 k-2 i a V_0\right) M_{k, \mu}\left(2 ia V_0 \right)-(1+2 k+2 \mu) M_{1+k}, \mu\left(2 i a V_0\right)=0.
\end{equation}

The equation \eqref{energy} represents the energy eigenvalue of the bound states for the DKP equation, which matches with the energy equation for the Klein--Gordon equation for a cusp potential\cite{villalba2006bound}.

To evidence the existence of the bound states for this potential and taking into account the energy equation Eq. \eqref{energy}, we set the parameters $a=0.5$.

\begin{figure}[th]
\begin{center}
\includegraphics[scale=0.6]{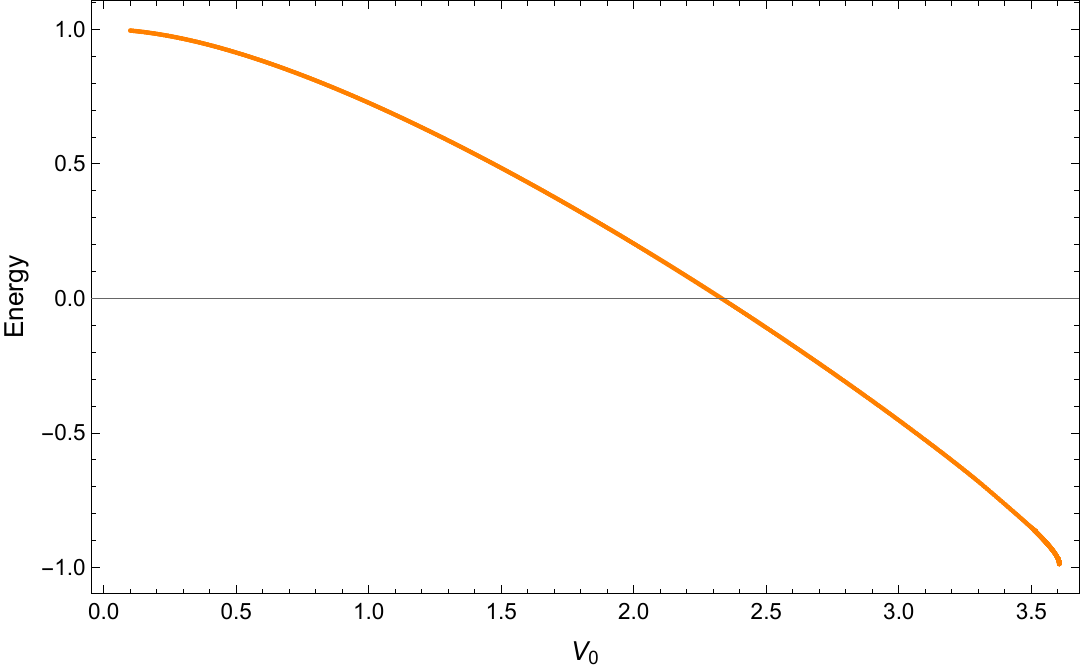}
\caption{Energy of the lowest bound--state spectrum for $a = 0.5$.}
\label{points}
\end{center}
\end{figure}

\begin{figure}[th]
\begin{center}
\includegraphics[scale=0.7]{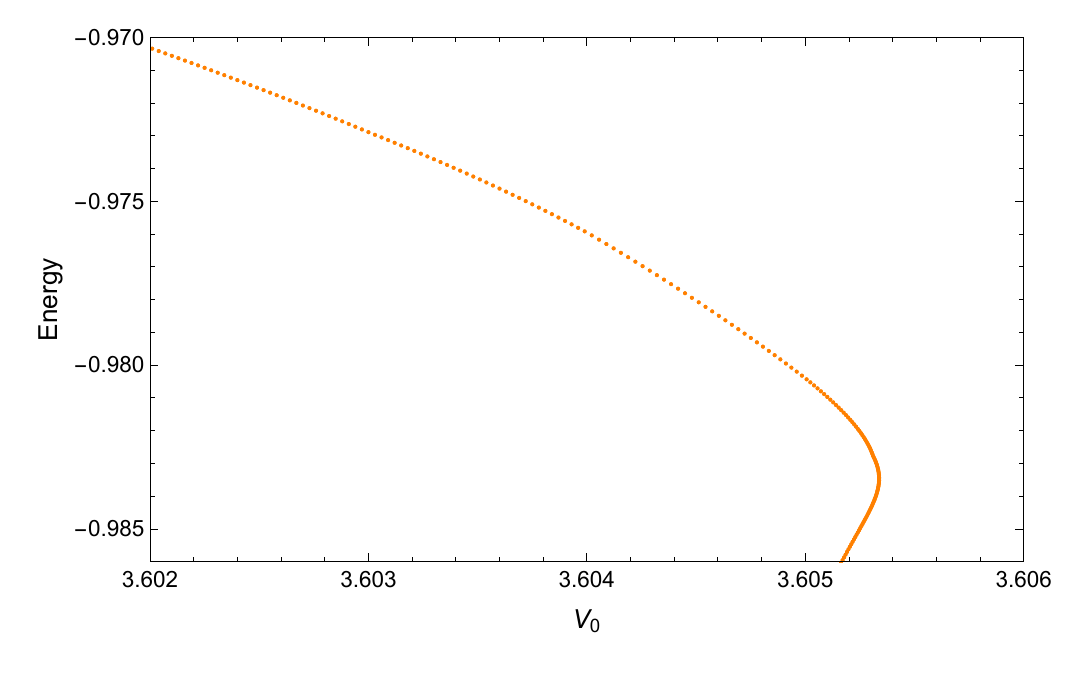}
\caption{Inset is an enlargement of the critical area, showing a turning point that indicates the particle and antiparticle state. The critical value for $V_0\approx 3.60534$ corresponds to $E \approx -0.98347$. Energy is given in units of the rest energy $mc^2$.}
\label{zoom}
\end{center}
\end{figure}

From Fig. \ref{zoom}, it is possible to see that there is a turning point at $V_0\approx 3.60534$ which represents that the potential supports a bound state with the specified conditions. Practically, it suggests that we found the condition under the pair creation happens. In other words, we have a state where it is dominated by the creation of particles and antiparticles\cite{reinhardt1977quantum,greiner2000relativistic,wachter2011relativistic}, which annihilate themselves due to the positive and negative energies.


\section{Conclusions}	
\label{sec_conclusion}

In this paper, we have solved the relativistic DKP equation for the cusp potential barrier within the framework of spin--one taking into account the equivalence with spin--zero particles in the DKP theory for $(1+1)$ dimensions. Our calculations involved determining the wave functions associated with both scattering and bound--states of the potential through the use of Whittaker functions. Furthermore, we have computed the transmission and reflection coefficients of the potential in great detail by analyzing the continuity and asymptotic properties of the wave functions. Moreover, we have shown that the energy equation related to the bound states in the DKP equation is the same result of a Klein--Gordon equation in a cusp potential well\cite{villalba2006bound}.



\end{document}